\documentclass[]{pasj02}
\usepackage{subcaption} 
\usepackage{ulem}

\jyear{2026}
\Received{}
\Accepted{}

\begin{document} 

\title{Progenitor and Explosion Mechanism of 3C~397 Indicated from XRISM High-resolution Spectroscopy of Fe-group Elements}

\author{
 Hiroshi \textsc{Nakajima},\altaffilmark{1}\altemailmark \email{hiroshi@kanto-gakuin.ac.jp} 
 Yuken \textsc{Ohshiro},\altaffilmark{2}\altemailmark \email{yuken.ohshiro@riken.jp}
 Shing-Chi \textsc{Leung},\altaffilmark{3,4}
 Koji \textsc{Mori},\altaffilmark{5}
 Yoshiaki \textsc{Kanemaru}\altaffilmark{6}
 Yoshitomo \textsc{Maeda}\altaffilmark{6}
 Hiroya \textsc{Yamaguchi}\altaffilmark{6}
 Masayoshi \textsc{Nobukawa}\altaffilmark{7}
 Kumiko \textsc{Nobukawa}\altaffilmark{8}
 Manabu \textsc{Ishida}\altaffilmark{6} and
 Yuichiro \textsc{Ezoe}\altaffilmark{9}
}
\altaffiltext{1}{College of Science and Engineering, Kanto Gakuinn University, Kanagawa 236-8501, Japan}
\altaffiltext{2}{RIKEN Pioneering Research Institute (PRI), Saitama 351-0198, Japan}
\altaffiltext{3}{Department of Physics, SUNY Polytechnic Institute, Utica, NY 13502, USA}
\altaffiltext{4}{Kavli Institute for the Physics and Mathematics of the Universe (WPI), The University of Tokyo, Chiba 277-8583, Japan}
\altaffiltext{5}{Faculty of Engineering, University of Miyazaki, Miyazaki 889-2192, Japan}
\altaffiltext{6}{Institute of Space and Astronautical Science (ISAS), Japan Aerospace Exploration Agency (JAXA), Kanagawa 252-5210,
Japan}
\altaffiltext{7}{Faculty of Education, Nara University of Education, Nara 630-8528, Japan}
\altaffiltext{8}{Department of Physics, Kindai University, Osaka 577-8502, Japan}
\altaffiltext{9}{Department of Physics, Tokyo Metropolitan University, Tokyo 192-0397, Japan}



\KeyWords{ISM: individual objects (3C~397) --- ISM: supernova remnants --- white dwarfs --- supernovae: individual (3C~397)}  

\maketitle

\begin{abstract}

We present spatially resolved X-ray spectroscopy of the Type Ia supernova remnant 3C~397, one of the most promising candidates to have originated from a white dwarf with a mass close to the Chandrasekhar limit.
The remnant was observed during the performance verification phase of the X-ray Imaging and Spectroscopy Mission (XRISM), with the field of view of the Resolve calorimeter array positioned on the eastern half of the remnant.
We divide the Resolve field of view into southeastern and northeastern regions and extract spectra from each region for both Resolve and Xtend.
The Resolve spectra are characterized by narrow K-shell emission lines from intermediate-mass elements (IMEs) such as Si, S, Ar, and Ca, and by broader K-shell emission lines from iron-group elements (IGEs) such as Cr, Mn, Fe, and Ni.
K-shell emission lines from Ti and Cr are also detected, and are found to be locally enhanced in the southeastern region, as reported in previous observations.
We model the Resolve and Xtend spectra simultaneously by a non-equilibrium ionization (NEI)
plasma model with multiple temperatures and ionization states.
It is found that the IMEs are represented with a low temperature component, while
the Ti and IGEs are in the multiple plasma states with relatively high temperatures and low
ionization states. The Ti and IGEs are in common plasma states with respect to
electron temperature, ionization state, and line-of-sight velocity dispersion, suggesting
that these elements originate from a common nucleosynthesis regime in the explosion.
The observed mass ratios of Ti/Fe and Cr/Fe in the southeastern region
can only be explained by nucleosynthesis in a neutron-rich environment produced
by electron-capture reactions in the innermost layers of the exploding white dwarf.
By comparing the observed mass ratios with those predicted by nucleosynthesis models, we constrain the central density to be $\gtrsim 4.0 \times 10^9~{\rm g~cm^{-3}}$ for the deflagration-to-detonation transition model and $\gtrsim 6.0 \times 10^9~{\rm g~cm^{-3}}$ for the pure turbulent deflagration model.
On the other hand, the observed Ni/Fe mass ratio is globally enhanced across the field of view, suggesting an additional neutronization mechanism beyond electron-capture reactions, such as a higher progenitor metallicity.
\end{abstract}


\section{Introduction}

Type Ia supernovae (SNe Ia), the thermonuclear explosions
of electron-degenerate carbon-oxygen white dwarfs (C+O WDs)
in a binary system \citep{Hoyle60},
contribute to various astrophysical fields such as cosmology
\citep{Riess16, Riess18}, galactic nucleosynthesis
(\cite{Matteucci21} and references therein),
and galaxy formation \citep{Gandhi22}.
In particular, a significant part of iron-group elements (IGEs) is
provided via SNe Ia
\citep{Kobayashi09, Matteucci12, Maoz17, Nomoto18, Kobayashi_2020}.
Although SNe Ia are thus of great cosmological and astrophysical importance,
their progenitor systems and explosion mechanism remain elusive
(\cite{Hillebrandt2000, Hillebrandt13, Maoz14, Ruiter25} and references therein).

Among the possible pathways, deflagration or delayed detonation
of a near-Chandrasekhar mass (near-$M_{\rm Ch}$) WD in a
single-degenerate system \citep{Whelan73} or 
in a double degenerate system \citep{Iben84}
certainly contributes to observational results.
For example, \citet{Mazzali07} argue for the near-$M_{\rm Ch}$ WDs based on
the velocity distributions of Si for nearby SNe Ia.
Another pathway is the so-called sub-$M_{\rm Ch}$ explosion, in
which there are two major channels; one is the violent mergers of
WD binaries \citep{Pakmor12} and the other one is accretion of
He from a companion onto the WD that triggers a double detonation \citep{Nomoto82, Goldstein18}.
It is suggested that the majority of SNe Ia is from sub-$M_{\rm Ch}$ explosion
by $\rm Ni~II/Fe~II$ ratios of SNe Ia spectra \citep{Flors20}.

To date, it has been shown that both pathways are indispensable
for explaining the galactic nucleosynthesis.
Abundances of Mn \citep{Seitenzahl13, Eitner20}
and Ni \citep{Palla21} of nearby stars require both
near-$M_{\rm Ch}$ WD and sub-$M_{\rm Ch}$ WD
explosions in roughly equal proportions.
High-resolution X-ray spectrum of the Perseus cluster also enabled
assessing the contributions of SN types; The mixture of
near-$M_{\rm Ch}$ and sub-$M_{\rm Ch}$ explosions reproduces
the abundance pattern of IGEs \citep{Hitomi17a}.
To refine population studies of SN Ia progenitor pathways,
it is crucial to measure the properties of individual SNe Ia,
such as their ejecta mass, progenitor metallicity, and evolution into supernova remnants.

Because the C+O WDs are totally degenerated,
the mass of the progenitor ($M_{\rm WD}$) and its central density
($\rho_c$) are directly correlated \citep{Chandrasekhar39}.
When the $M_{\rm WD}$ is close to $M_{\rm Ch}$ and SN occurs,
$\rho_c$ reaches $\ge 10^9 {\rm g/cm^3}$. 
Such an extremely dense condition induces efficient electron capture
reactions, reducing the number of free electrons. Then
the neutronized IGEs
such as $^{48-50}{\rm Ti}$, $^{50-54}{\rm Cr}$, $^{55}{\rm Mn}$,
$^{54,56,58}{\rm Fe}$, $^{57,58,62}{\rm Ni}$
are produced \citep{Woosley97,Iwamoto99,Brachwitz00}
in neutron-rich nuclear statistical equilibrium (n-rich NSE).
Accordingly, the WD explosion mechanism strongly influences
the abundance ratios among IGEs in a supernova remnant (SNR).
In such a circumstance, spatially resolved
X-ray spectroscopy of SNRs is the
most powerful approach because the IGE ratios and their distributions
can directly be obtained. 

Recently launched XRISM \citep{Tashiro25}
has opened a new era of X-ray astronomy by the combination of
extremely high energy resolution by Resolve \citep{Kelley25, Ishisaki25}
and large effective area by Xtend \citep{Noda25, Uchida25}.
This opportunity motivates us to observe 
3C~397, one of the best targets for investigating the
progenitor and explosion mechanisms of near-$M_{\rm Ch}$ SNe Ia.
It is a middle-aged SNR 
located near the Galactic plane \citep{Chen99, Safi-Harb05}
at a distance of 8~kpc \citep{Leahy16, Ito25}.
One of the most remarkable observational features of this SNR is the detection
of strong K-shell emission lines of IGEs in the X-ray band \citep{Yamaguchi15, Ohshiro21}
with Suzaku \citep{Mitsuda07, Koyama07} and with XMM-Newton \citep{Jansen01}.
In particular, \citet{Ohshiro21} reported the detection of the K-shell emission
lines of stable Ti. They found that the high mass ratios of Ti and IGEs relative to Fe
can only be explained by a near-$M_{\rm Ch}$ explosion with 
a high $\rho_c$.
However, limited photon statistics around the energy band of IGEs hindered
us from precisely estimating $\rho_c$.
In this paper we report a new insight of Ti and IGEs ejecta of 3C~397
realized with a long exposure data of XRISM.

\section{Observation}\label{sec:2}

The first observation of 3C~397 by XRISM
was carried out during its performance verification (PV)
phase as summarized in
Table~\ref{tab:obs}.
Although the exposure periods were divided only
by programmatic circumstances, the pointing directions and roll
angles were set to be the same among the three exposures.
As a result, the field of view (FoV) differences
among the observations were smaller than one pixel of Xtend
that corresponds to $\timeform{1".77}$.
The brightness of the remnant is moderate when taking into
account the operating mode of Resolve and Xtend. Therefore,
the filter setting of Resolve was open, and
Xtend was operated with the ``Full
Window + No Burst'' mode throughout the observation.
Note that the gate valve of Resolve \citep{Midooka21}
was still closed during the observation and hence
the lower edge of its effective energy band was approximately
1.8~keV.

We reprocessed the unfiltered data with the XRISM software
tools V002 \citep{Doyle22} included in HEASOFT version 6.35.2.
We used the calibration files for Resolve and
Xtend, published in October 2025.
After data reduction, the total effective
exposure times of Resolve and Xtend are 477.2 and
465.6~ks, respectively.
Several reports of high solar activity during the XRISM PV phase raised concerns about possible spectral contamination in the soft X-ray band \citep{Suzuki25}.
We examined the Resolve and Xtend light curves extracted from the region corresponding to the Resolve FoV in the 2.0--10.0~keV band and above 10.0~keV, and confirmed that neither showed significant temporal variability.
Therefore, we summed the data of three exposures
for each instrument.

\begin{figure}
 \begin{center}
  \includegraphics[width=7cm]{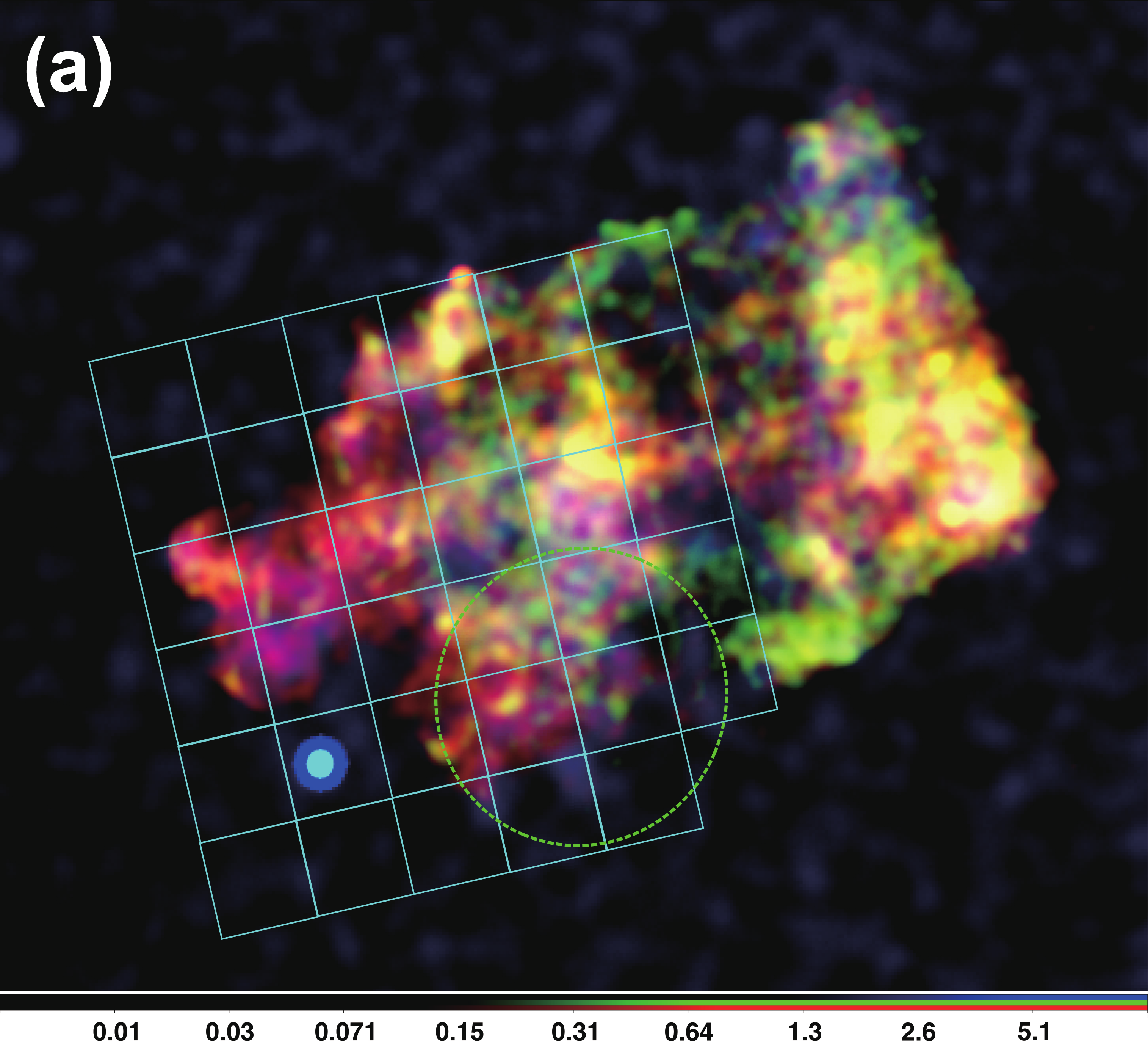} \\
  \includegraphics[width=7cm]{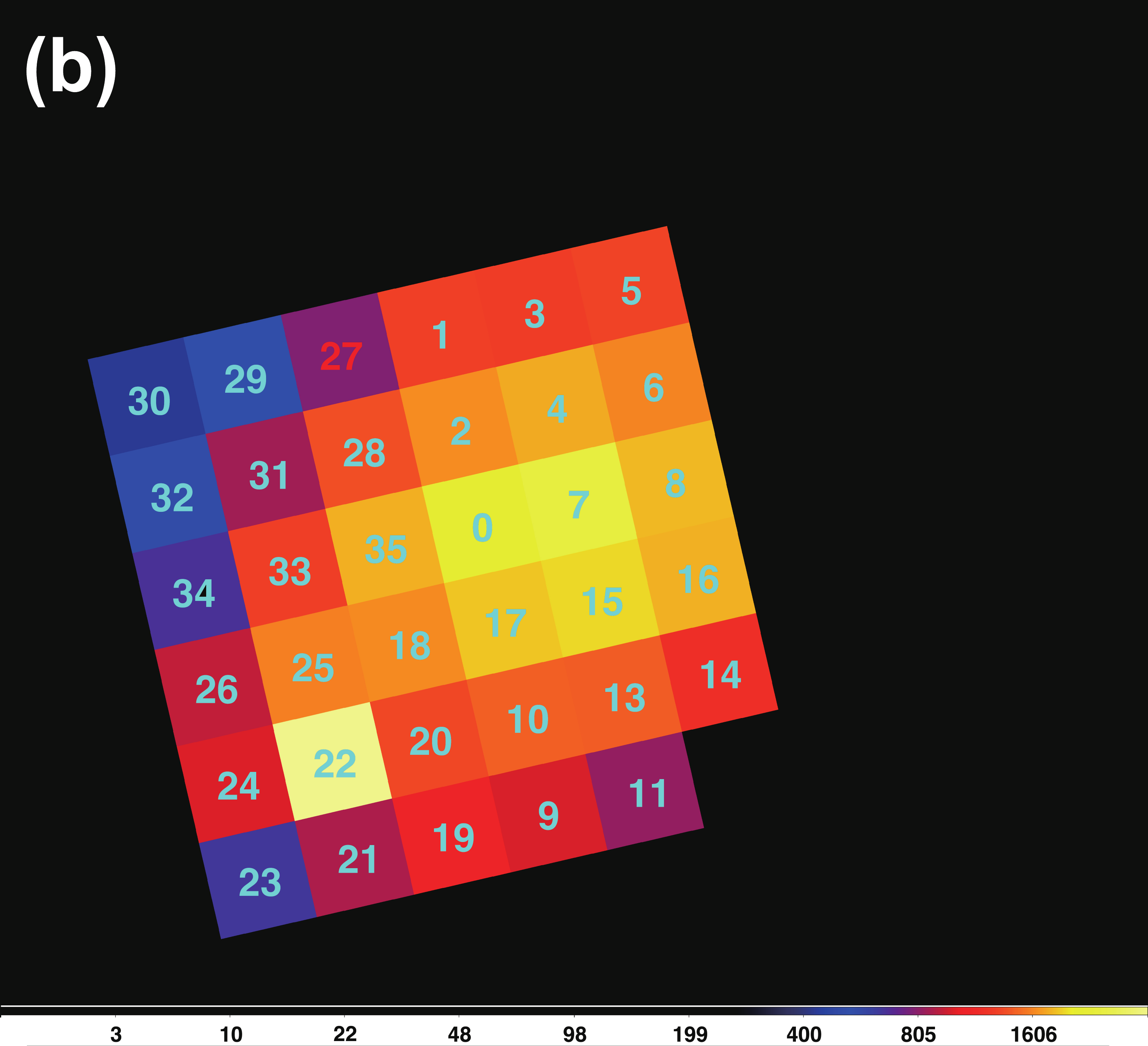} 
  \includegraphics[width=7cm]{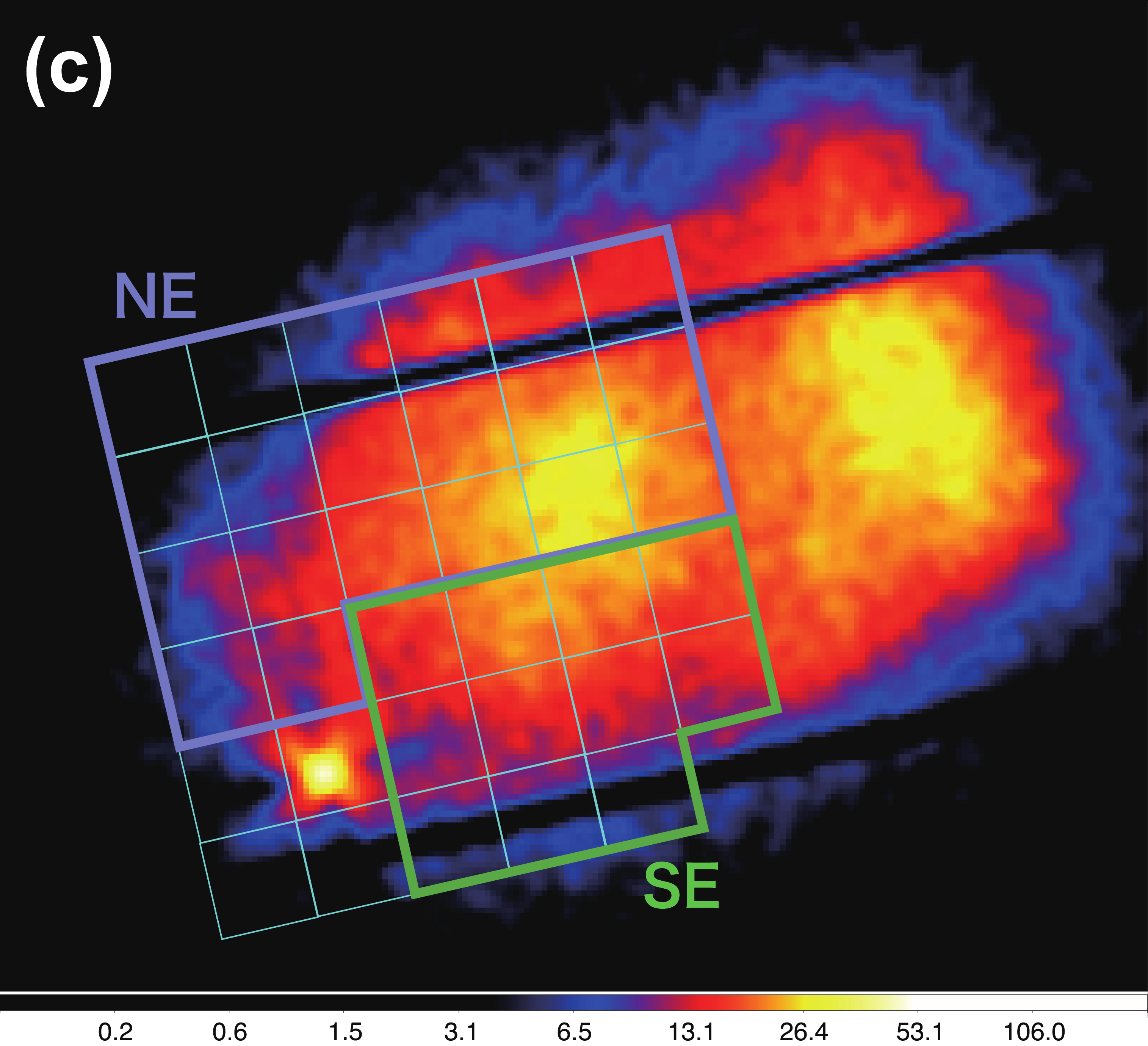} 
 \end{center}
\caption{(a) Three-color image of 3C~397 by Chandra with the
energy range from 0.5--10.0~keV. Resolve pixel array (cyan squares)
and the Cr-rich region identified by \citet{Ohshiro21} (green
dashed ellipse) are overlaid.
(b) Resolve count map in the energy range
from 2.0--10.0~keV with the pixel IDs.
The calibration pixel at the bottom right
corner of the array is not exposed.
(c) Xtend count map with the same energy band as that of the Resolve image.
The photon-extracted regions for
spectral analyses are defined as polygons with
green (southeastern region) and blue (northeastern region).
Two line-shaped missing areas across the remnant
designate the charge injection lines,
which are taken into account in the calculation of the effective area. 
{Alt text: Three images showing the X-ray count maps obtained by Chandra, XRISM/Resolve, and XRISM/Xtend.}
}\label{fig:image}
\end{figure}

The spatial distribution of the remnant is approximately
$3'\times5'$ that is elongated from east to west,
as shown in the three-color image of Chandra
in figure~\ref{fig:image} (a).
The center in the east-west direction is relatively dim,
and there are two bright lobes in the west and east.
In particular, the X-ray emission is enhanced around
the inner edge of the eastern lobe.
The FoV of Resolve in the PV observation
covers the eastern lobe of the remnant
and is indicated by the squares in figure~\ref{fig:image} (a).
It is set to include the Cr-rich region reported by \citet{Ohshiro21}
that is shown with an ellipse.
The count maps of Resolve and Xtend are shown in figure~\ref{fig:image}
(b) and (c), respectively.
The point source CXO~J190741.2+070650, located to the southeast of the remnant and identified in a 2001 Chandra observation \citep{Safi-Harb05}, is clearly detected in both panels.
Because of the relatively limited angular resolution of XRISM, contamination from the point source should be evaluated in the following analysis.
Since it is not associated with the remnant \citep{Safi-Harb05}, a detailed analysis of the point source and a discussion of its nature will be presented in a separate paper (Ohshiro et al., in prep.).

\begin{table}
  \tbl{Pointing direction and exposure time.}{%
  \begin{tabular}{cccccc}
      \hline
      Obs. ID & R.A. & Dec. & Roll angle~(deg) & \multicolumn{2}{c}{Exposure~(ks)\footnotemark[$*$]}  \\ 
       & \multicolumn{2}{c}{(J2000.0)} & & Resolve & Xtend  \\ 
      \hline
      300013010 & $\timeform{286D.907}$ & $\timeform{7D.130}$ & 103.00 & 200.16 & 195.23 \\
      300013020 & $\timeform{286D.907}$ & $\timeform{7D.130}$ & 103.00 & 176.23 & 175.92 \\
      300013030 & $\timeform{286D.907}$ & $\timeform{7D.130}$ & 103.00 & 100.80 & 94.41 \\
      \hline
    \end{tabular}}\label{tab:obs}
\begin{tabnote}
\footnotemark[$*$] Effective exposure times after the data screening.  \\ 
\end{tabnote}
\end{table}

\section{Analysis}\label{sec:ana}

\begin{figure}
 \begin{center}
  \includegraphics[width=\columnwidth]{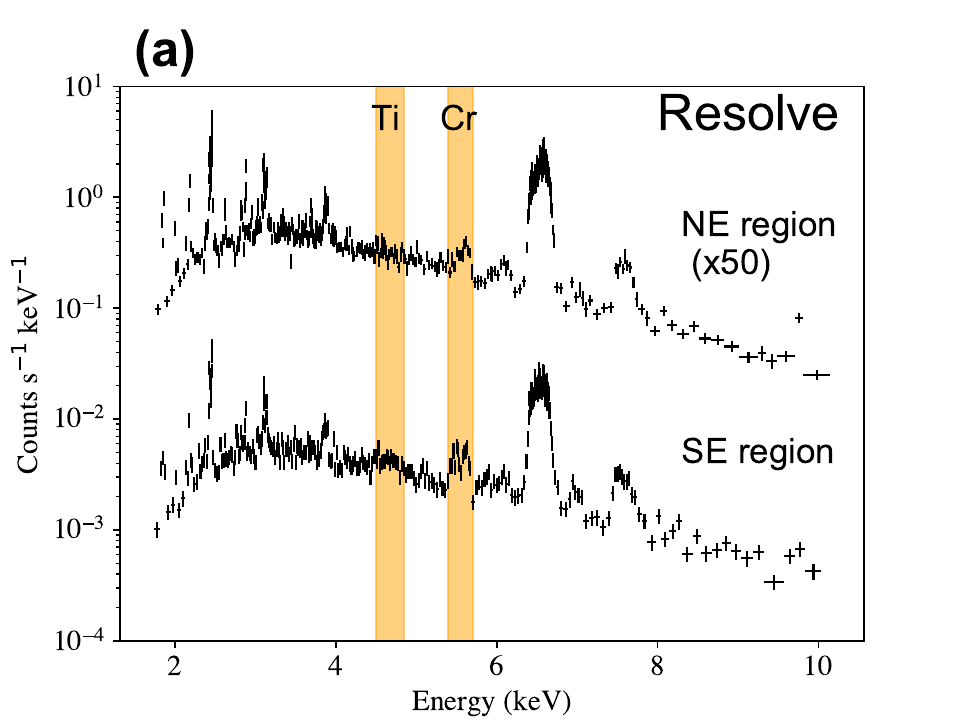} 
  \includegraphics[width=\columnwidth]{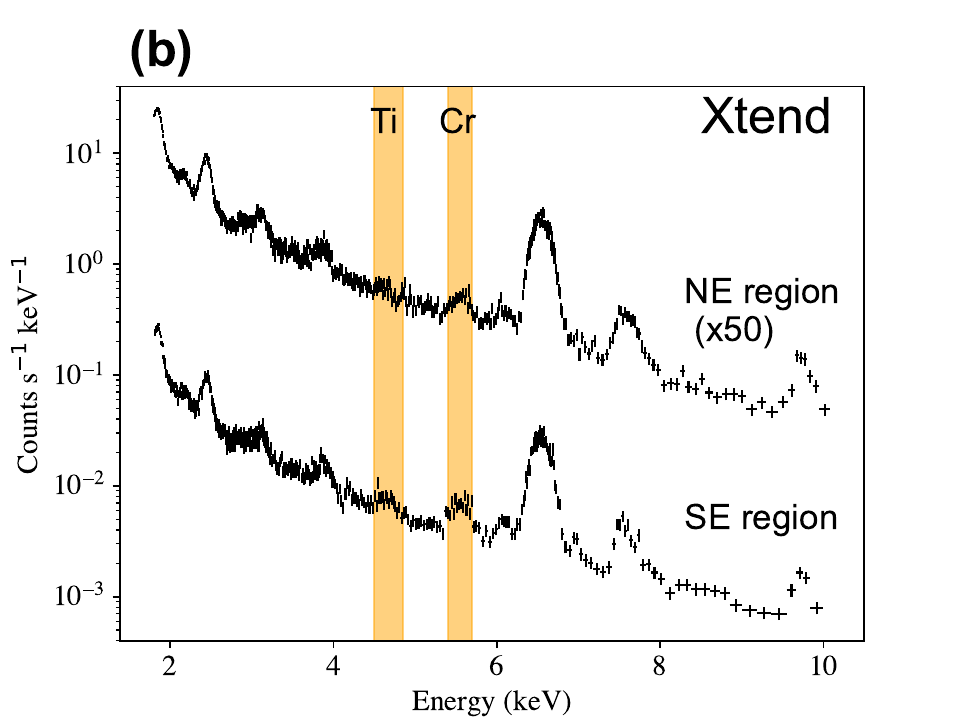} 
 \end{center}
\caption{(a) Resolve spectra extracted from SE and NE regions.
The latter is multiplied by 50 for visibility.
K-shell lines from Ti and Cr, the fluxes of which exhibit the most apparent
differences between the regions, are highlighted in orange.
Note that a line around 9.7~keV is an instrumental line as modeled
in fugure~\ref{fig:TiRichspectra_SE} and \ref{fig:TiNonRichspectra_NE}.
(b) Same as the top panel, but for Xtend spectra.
{Alt text: Two line graphs showing the X-ray spectra obtained by Resolve and Xtend from the NE and SE regions.}
}\label{fig:SE-NE-spectra}
\end{figure}

According to imaging analyses performed by \citet{Ohshiro21},
the southeastern part of the remnant exhibits relatively
higher flux from Cr than other regions. Based on their results,
we define two regions split by the pixel boundaries of Resolve
as shown in figure~\ref{fig:image} (c).
The Southeastern (SE) region includes 11 pixels enclosed with a green polygon that
includes the Cr-rich region identified with XMM-Newton \citep{Ohshiro21}.
For comparison, we also investigate the northeastern (NE) region,
which consists of the remaining Resolve pixels enclosed within a blue polygon.
Note that pixel 27 is excluded due to its deviation in gain performance,
and pixel 22 is also excluded because it includes
the point source.

\begin{figure*}
 \begin{center}
  \includegraphics[width=\textwidth]{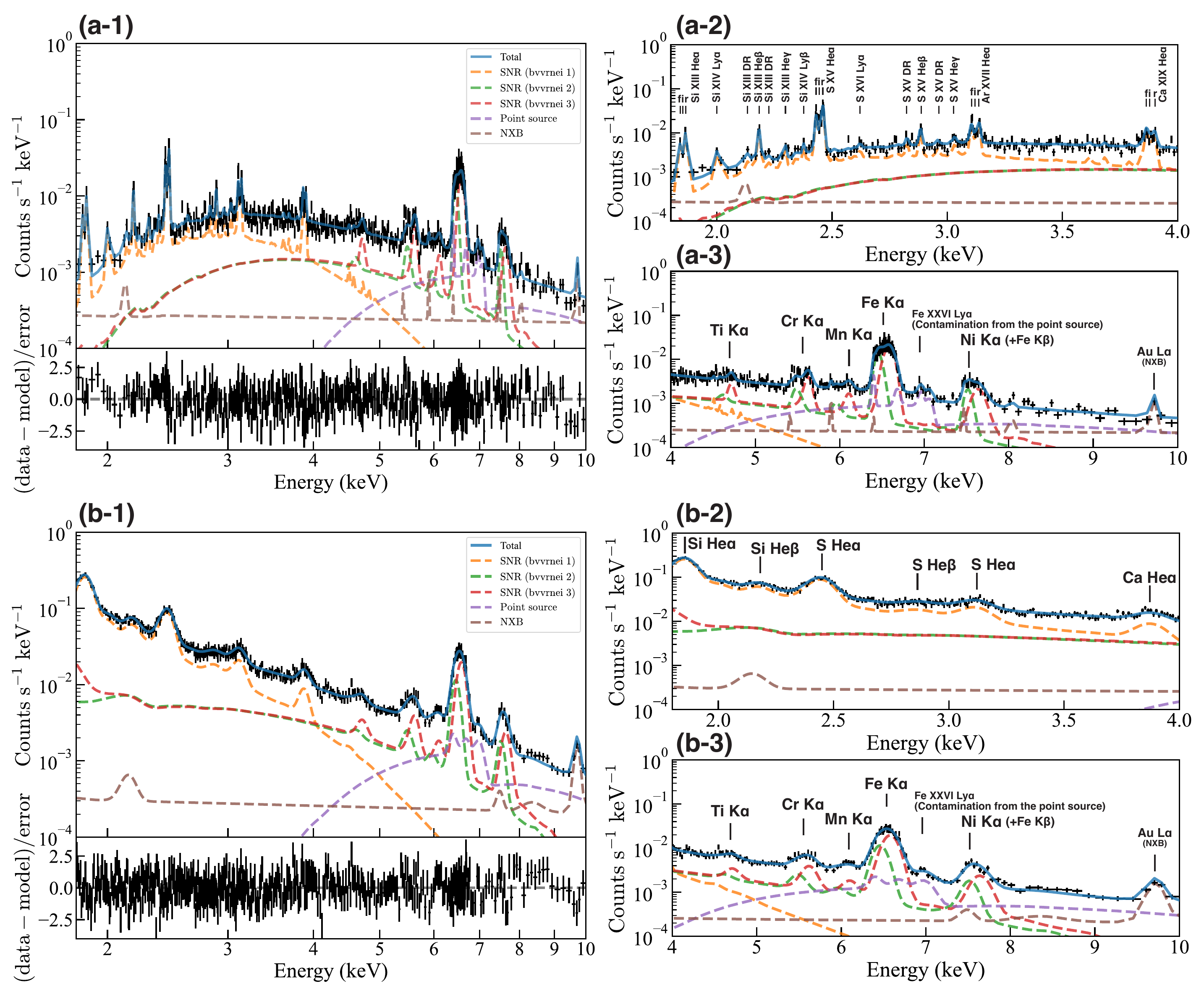} 
 \end{center}
\caption{(a-1) Resolve spectrum extracted from the SE region and the best-fit model with \texttt{xspec} in the 1.8--10.0 keV band. (a-2) and (a-3) show the same spectrum in the 1.8–4.0 keV and 4.0–10.0 keV bands, respectively, with line identifications.
(b-1), (b-2), and (b-3) are the same as (a-1), (a-2), and (a-3), respectively, but for Xtend spectrum.
{Alt text: Six line graphs showing the spectra extracted from the SE region with fitting curves.}}\label{fig:TiRichspectra_SE}
\end{figure*}

\begin{figure*}
 \begin{center}
  \includegraphics[width=\textwidth]{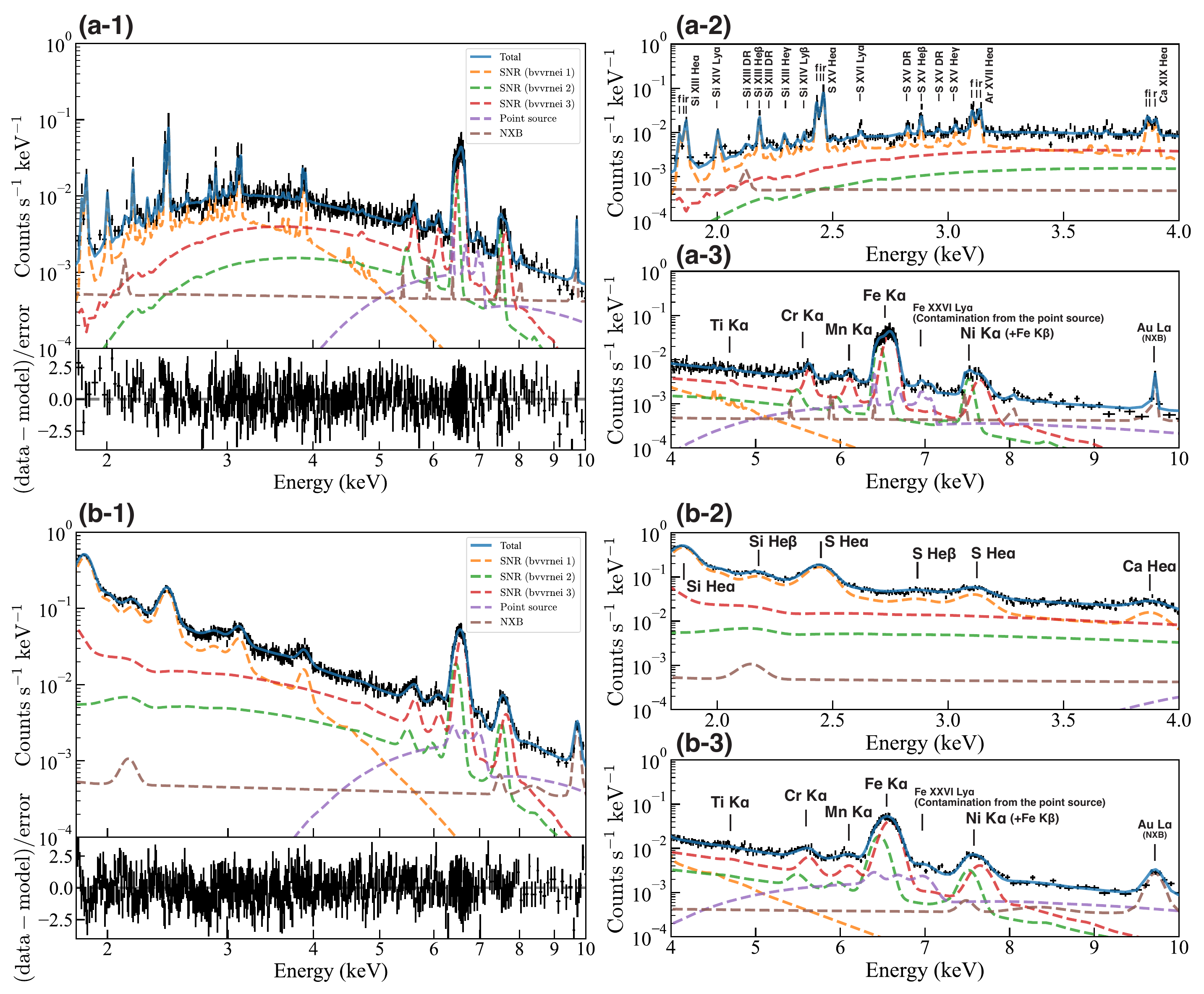} 
 \end{center}
\caption{Same as figure~\ref{fig:TiRichspectra_SE}, but for the spectra extracted from the NE region.
{Alt text: Six line graphs showing the spectra extracted from the NE region with fitting curves.}}\label{fig:TiNonRichspectra_NE}
\end{figure*}

The extracted spectra of Resolve and Xtend from the SE and
NE regions are shown in figure~\ref{fig:SE-NE-spectra}.
Because of the limited effective area of Resolve and
the ongoing calibration of Xtend at the soft energy band,
we focus on spectra above 1.8~keV for both instruments.
The spectra exhibit strong emission lines from intermediate
mass elements (IME: Si, S, Ar, Ca, Ti) and IGEs in all spectra.
The remarkable differences between the two regions
are the line flux and the line profile of Ti and Cr.
The K-shell Ti line is dim at the NE region but bright
in the SE region as confirmed by both instruments,
which indicates its localization inside the remnant. 
The difference can be better identified in Xtend spectra,
thanks to its large effective area, low background level,
and moderate energy resolution of CCD detectors.
The Cr line in the Resolve spectrum is split into two for the SE region,
which indicates the existence of at least two ionization states or 
line-of-sight velocities. Although there is a strong
emission line from Cr also in the NE region, the line
fluxes differ significantly between the two regions.
Hence, our result does not contradict the non-uniformity
of Cr reported by \citet{Ohshiro21}.

The details of the spectra in the SE and NE region
are illustrated for narrow energy bands
in figure~\ref{fig:TiRichspectra_SE} and
figure~\ref{fig:TiNonRichspectra_NE}, respectively.
Thanks to the high spectral resolution power,
Resolve spectra clearly show narrow emission lines from
IMEs below 4~keV.
Especially for S and Ar, triplet lines from He-like
ions are well resolved so that the line widths can be
constrained for the first time for this remnant.
He~$\beta$, He~$\gamma$, and dielectronic 
recombination (DR) satellite lines
can also be observed for those ions, enabling us
to constrain the plasma temperature and ionization
parameter more precisely than in past observations.
Contrastingly, emission lines from Ti and IGEs
are relatively broad, which strongly indicates that
the two groups of elements
are in different dynamical states.
Especially, the asymmetrical profile can
be seen for He$\alpha$ from Fe.
The above differences between IMEs and IGEs clearly suggests
that they originate from different plasma states; IMEs are
near ionization equilibrium, while Ti and IGEs are still
in an ionizing state.
All of the characteristics are identified for the first time
by Resolve and have not been observed in any of the CCD data
ever obtained. 

To characterize the spectra, the response matrix files
were generated by \texttt{rslrmf} for Resolve with an option
that incorporates the escape peaks and the Si K$\rm \alpha$ lines,
and by \texttt{xtdrmf} for Xtend.
The energy-dependent effective areas (the auxiliary files)
are created with \texttt{xaarfgen} that runs a ray-tracing simulation
with an input of source image. In this work, we adopted the Chandra
image of the remnant in a energy band of 2-10~keV.
Further spectral analyses are performed with
\texttt{xspec} version 12.15.0d \citep{Arnaud96} and
\texttt{SPEX} version 3.08.02 \citep{Kaastra96} to
estimate the systematic uncertainty of the derived
best-fit parameters. Regarding the former case,
\texttt{AtomDB} \citep{Smith01, Foster12}
version 3.1.2 is adopted.
The applied set of response and auxiliary files for each model
component is the same between the analyses using \texttt{AtomDB} and \texttt{SPEX}.

Background subtraction is not performed in the spectra
in figure~\ref{fig:TiRichspectra_SE} and
figure~\ref{fig:TiNonRichspectra_NE}, which requires us to
evaluate the contribution from
the non-X-ray background (NXB)
and any objects other than the remnant.
All spectra exhibit significant emission lines around 7.0~keV
that are presumably from H-like Fe ions. This indicates
the possible contamination of the spectra
from CXO~J190741.2+070650 due to the extension of the point spread
function of the X-ray mirror assembly.
Then we incorporated the point source component
with the spectral parameters that are derived by Ohshiro et al. in prep.
The effective area for the point source component
is calculated individually by \texttt{xaarfgen}
using the source position as input.
NXB level is sufficiently low
compared with that of the remnant \citep{Kilbourne18}
below the energy band of IGEs.
However, there is an appreciable
emission line above 9~keV
(neutral Au L$\alpha$) that originates within
the satellte, which indicates
the significant contribution by NXB above the IGEs band.
Hence, we included the phenomenological model for each instrument.
Relative intensities among continuum power-law and
several Gaussian lines in NXB spectra are fixed to
those of the model released from the XRISM team
\footnote{https:\slash\slash{}heasarc.gsfc.nasa.gov\slash{}docs\slash{}xrism\slash{}analysis\slash{}nxb}.
Only its normalization
is set to be free for each instrument
by applying a constant factor. Diagonal response matrices
are used for NXB components of both instruments.
Sky foreground and background components are omitted for the
reason that their fluxes are primarily in the soft energy band
below 1~keV and can be ignored compared with the source emissions.

Taking into account the above attributes of the spectra,
we adopted a model of optically-thin thermal plasma
in a non-equilibrium of ionization (NEI) state
with a single temperature, ionization
parameter ($\tau$), and velocity dispersion ($v_{\rm rms}$).
The NEI plasma model is applied with \texttt{bvvnei} for \texttt{AtomDB}
and \texttt{Neij} for \texttt{SPEX} \citep{Kaastra93}.
In addition to the primary emission lines, many satellite
lines from low-ionized ions are present in the case of
low-ionization-parameter plasma. Those lines are set to be broad as
those of primary lines in the spectral fit using \texttt{xspec}
(\texttt{APECBROADPSEUDO} is activated).
We first applied a single NEI component to the spectra and found
that the model can not reproduce the different emission line profiles
of IMEs and IGEs. Next, we tried two NEI components with different
temperatures, $\tau$, and $v_{\rm rms}$.
It was found that the model adequately reproduced
the emission lines from IMEs with one component.
However, the other component cannot solely reproduce
spectra above 4~keV. This is reasonable because the line
profile of Fe He$\alpha$ is asymmetric,
presumably due to the presence of multiple plasma
components with different temperatures or ionization parameters.
When we fit the line with two gaussian functions, the
difference of the mean values is approximately 150~eV
that corresponds to 6600~${\rm km/s}$.
The large velocity difference can be naturally
explained by receding and approaching ejecta components
moving nearly parallel to the line-of-sight.
This is reasonable if the extracting
regions are around the center of the explosion. However,
especially for the SE region, it covers only the edge of
the remnant rather than the center. Then the difference
in the line center is plausibly interpreted as due to
different temperatures and/or ionization parameter.
The asymmetrical profile of the emission lines can
also be seen for Cr, and there is no clear difference
among the elements in terms of the line profile. Hence,
the abundance pattern among the elements can be
assumed to be the same between the components.
Then we finally adopted the three
temperature components: one for low-temperature NEI
plasma that mainly reproduces the low-energy part of
the spectra including IME emissions, and the other two are
high-temperature components that explain most of the
emission from Ti and IGE.
The origin of the latter two
components is assumed to be the SN ejecta.
Therefore, the abundances of those elements are set
to be high enough
(in units of $10^8$ solar) as adopted in
\citet{Ohshiro21} in order to ignore the thermal
bremsstrahlung emission between other light elements
and electrons.

\begin{longtable}{ccccc}
  \caption{Best-fit plasma parameters in the model fitting and mass ratios.}\label{tab:bestfitparam}  
\hline\noalign{\vskip3pt} 
Region & \multicolumn{2}{c}{NE} & \multicolumn{2}{c}{SE}  \\   [2pt] 
\hline
Database     & ATOMDB & SPEX & ATOMDB & SPEX  \\   [2pt] 
Version    & 3.1.2 & 3.08.02 & 3.1.2 & 3.08.02  \\   [2pt] 
\hline\noalign{\vskip3pt} 
\endfirsthead      
\hline\noalign{\vskip3pt} 
Region & \multicolumn{2}{c}{NE} & \multicolumn{2}{c}{SE}  \\  [2pt] 
\hline\noalign{\vskip3pt} 
\endhead
\hline\noalign{\vskip3pt} 
\endfoot
\hline\noalign{\vskip3pt} 
\multicolumn{5}{@{}l@{}}{\hbox to0pt{\parbox{160mm}{
\footnotesize
\hangindent6pt\noindent
\hbox to6pt{\footnotemark[$*$]\hss}\unskip%
   Model components correspond to those noted in figure~\ref{fig:TiRichspectra_SE} and
   figure~\ref{fig:TiNonRichspectra_NE}.\\
\hbox to6pt{\footnotemark[$**$]\hss}\unskip%
\ Emission measures are derived assuming a distance of 8~kpc \citep{Leahy16,Ito25}.
}\hss}}
\endlastfoot 
\multicolumn{5}{c}{Absorption} \\
    &  \texttt{TBabs} & \texttt{Absm} & \texttt{TBab} & \texttt{Absm}  \\
    \cline{2-5} 
$N_{\rm H}~(10^{22}{\rm cm^{-2}})$ & \multicolumn{4}{c}{3.0(fix)}  \\

\hline
\multicolumn{5}{c}{Low temperature component} \\[1pt]
    & \texttt{bvvrnei 1}\footnotemark[$*$] &  \texttt{Neij}  & \texttt{bvvrnei 1}\footnotemark[$*$] & \texttt{Neij} \\ 
    \cline{2-5}
  $kT_e$~(keV)& $0.69^{+0.01}_{-0.01}$ & $0.72^{+0.01}_{<-0.01}$ &  $0.66^{+0.01}_{-0.01}$ & $0.68^{+0.01}_{<-0.01}$\\
  $\tau~(10^{12} {\rm cm^{-3}s})$ & $1.19^{+0.51}_{-0.13}$ & $>1.23$ & $>1.04$ & $>2.95$  \\
  $\sigma_v~({\rm km~s^{-1}}$) & $521^{+27}_{-26}$ & $511^{+27}_{-26}$ & $474^{+48}_{-46}$ & $457^{+48}_{-45}$  \\
  $n_e n_{\rm H}V~(10^{57} {\rm cm^{-3}})$\footnotemark[$**$] & $3.41^{+0.16}_{-0.10}$ & $25.9^{+0.1}_{-0.1}$ & $2.66^{+0.12}_{-0.13}$  & $25.9^{+0.1}_{-0.1}$   \\
  Si~(solar) & $1.75^{+0.07}_{-0.07}$ & $2.15^{+0.02}_{-0.05}$ & $1.44^{+0.07}_{-0.07}$ & $1.30^{+0.02}_{-0.01}$ \\
  S~(solar)  & $1.76^{+0.08}_{-0.05}$ & $2.16^{+0.03}_{-0.06}$ & $1.46^{+0.06}_{-0.05}$ & $1.40^{+0.04}_{-0.03}$ \\
  Ar~(solar) & $2.18^{+0.15}_{-0.10}$ & $2.66^{+0.12}_{-0.11}$ & $1.69^{+0.12}_{-0.12}$ & $1.68^{+0.11}_{-0.10}$ \\
  Ca~(solar) & $3.95^{+0.28}_{-0.32}$ & $4.48^{+0.25}_{-0.24}$ & $4.04^{+0.43}_{-0.31}$ & $3.60^{+0.28}_{-0.28}$ \vspace{1mm} \\
\hline
\multicolumn{5}{c}{High temperature components} \\[1pt]
   & \texttt{bvvrnei 2}\footnotemark[$*$] &  \texttt{Neij}  & \texttt{bvvrnei 2}\footnotemark[$*$] & \texttt{Neij} \\
    \cline{2-5}
  $kT_e$~(keV)     & $2.15^{+0.05}_{-0.08}$ & $1.96^{+0.09}_{-0.08}$ & $1.82^{+0.12}_{-0.09}$ & $1.97^{+0.01}_{-0.02}$  \\
  $\tau~(10^{10} {\rm cm^{-3}s})$ & $1.51^{+0.10}_{-0.09}$ & $1.65^{+0.15}_{-0.21}$ & $1.40^{+0.14}_{-0.13}$ & $0.53^{<0.01}_{-0.02}$  \\
  $\sigma_v~({\rm km~s^{-1}}$) & $1525^{+75}_{-74}$ & $1548^{+70}_{-70}$ & $1559^{+106}_{-107}$ & $1976^{+121}_{-120}$  \\
  $n_e n_{\rm H}V~(10^{49} {\rm cm^{-3}})$\footnotemark[$**$] & $4.40^{+1.32}_{-0.37}$ & $56.6^{+1.2}_{-11.6}$  & $5.02^{+1.22}_{-1.01}$ & $44.3^{+1.2}_{-1.0}$  \\
  Ti~($10^8$solar) & $0.64^{+0.38}_{-0.38}$ & $0.85^{+0.28}_{-0.27}$ & $4.64^{+0.73}_{-0.68}$ &  $3.65^{+0.53}_{-0.51}$\\
  Cr~($10^8$solar) & $3.22^{+0.22}_{-0.21}$ & $1.94^{+0.11}_{-0.11}$ & $5.68^{+0.39}_{-0.35}$ & $3.04^{+0.21}_{-0.20}$ \\
  Mn~($10^8$solar) & $4.88^{+0.47}_{-0.46}$ & $2.95^{+0.23}_{-0.22}$ & $5.11^{+0.77}_{-0.70}$ & $2.67^{+0.37}_{-0.34}$  \\
  Fe~($10^8$solar) & \multicolumn{4}{c}{1.0 (fix)}  \\
  Ni~($10^8$solar) & $10.32^{+0.42}_{-0.40}$ & $5.26^{+0.24}_{-0.24}$ & $10.65^{+0.59}_{-0.46}$ & $5.56^{+0.22}_{-0.20}$ \\
  & & & & \\
    & \texttt{bvvrnei 3}\footnotemark[$*$] &  \texttt{Neij}  & \texttt{bvvrnei 3}\footnotemark[$*$] & \texttt{Neij}\\
    \cline{2-5}
 $kT_e$~(keV)  & $1.76^{+0.05}_{-0.04}$  & $1.45^{+0.03}_{-0.03}$ & $1.99^{+0.20}_{-0.12}$ & $1.67^{+0.03}_{-0.04}$  \\
 $\tau~(10^{10} {\rm cm^{-3}s})$ & $9.50^{+0.36}_{-0.53}$ & $14.46^{+0.95}_{-0.15}$ & $7.37^{+0.68}_{-0.86}$ & $7.98^{+0.25}_{-0.24}$  \\
 $n_e n_{\rm H}V~(10^{49} {\rm cm^{-3}})$\footnotemark[$**$] & $5.41^{+0.29}_{-0.48}$ & $106^{+2}_{-8}$  & $2.19^{+0.41}_{-0.45}$  & $55.3^{+0.8}_{-0.6}$  \vspace{1mm} \\
\hline
\multicolumn{5}{c}{Statistics} \\[1pt]
 C-Statistis & 1459.41 & 1382.44 & 845.37 &  860.20 \\
 d.o.f. & 1172 & 1056 & 789 &  787 \\
\hline
\multicolumn{5}{c}{Mass ratios} \\[1pt]
 Ti/Fe~($10^{-3}$) & $1.15^{+0.68}_{-0.68}$ & $1.54^{+0.54}_{-0.50}$ & $8.32^{+1.31}_{-1.22}$ & $6.54^{+1.11}_{-0.95}$ \\
 Cr/Fe~($10^{-2}$) & $3.00^{+0.21}_{-0.20}$ & $1.82^{+0.11}_{-0.11}$ & $5.29^{+0.36}_{-0.33}$ & $2.83^{+0.18}_{-0.19}$ \\
 Mn/Fe~($10^{-2}$) & $2.51^{+0.24}_{-0.24}$ & $1.52^{+0.12}_{-0.13}$ & $2.63^{+0.40}_{-0.36}$ & $1.38^{+0.19}_{-0.19}$ \\
 Ni/Fe~($10^{-1}$) & $4.13^{+0.17}_{-0.16}$ & $2.10^{+0.09}_{-0.10}$ & $4.26^{+0.24}_{-0.18}$ & $2.22^{+0.09}_{-0.08}$ \\[1pt]
\hline
\end{longtable}

Best-estimated parameters and C-statistics \citep{Cash79}
obtained in the simultaneous
spectral fit of Resolve and Xtend are listed in
Table~\ref{tab:bestfitparam}.
As for the column density of the interstellar absorption,
we use the Tuebingen-Boulder interstellar medium absorption
(\texttt{TBabs}) model \citep{Wilms00} in \texttt{xspec}, while
the interstellar photoelectric absorption cross sections
(\texttt{Absm}) model \citep{Morrison83} is used for \texttt{SPEX}. In both cases,
the value of the column density is fixed to that
reported by \citet{Safi-Harb05}. 
The solar abundance table is derived
from values in \citet{Anders1989} and \citet{Lodders09} for
\texttt{xspec} and \texttt{SPEX}, respectively.
All the parameters of the SNR and contamination from the
point source are tied between the instruments. When Resolve
and Xtend spectra are individually investigated, the derived
flux coincides with each other. Hence, the constant factor between
the instruments is fixed to be one.
Low NXB level and best photon statistics ever obtained
make it possible for us to precisely estimate the abundance
of Ti and IGEs compared with other past observations, which is
important in the discussion below.
Note that the preshock temperature and Hydrogen density
defined in the \texttt{Neij} model in \texttt{SPEX} are set to
sufficiently low values of 2~eV and 1~$\rm{cm^{-3}}$,
respectively.

Mass ratios of Ti, Cr, Mn, and Ni to Fe are attached in
the bottom of Table~\ref{tab:bestfitparam}.
To derive these values, we adopt the following 
average atomic weights in the solar system ($m_{\rm Ti},
m_{\rm Cr},m_{\rm Mn},m_{\rm Fe},m_{\rm Ni}$ are 47.9, 52.0, 54.9,
55.8, and 58.7, respectively). The Solar abundance table is taken
from \citet{Anders1989}.
Ejecta in the SE region exhibits higher mass ratios
for almost all of the four elements. The most remarkable difference
between the two regions is the Ti/Fe ratio. Approximately one order
of magnitude higher value is obtained in the SE region,
which is the quantitative evidence of the Ti localization.

\section{Discussion}

\subsection{Constraints on the Progenitor Central Density}
\label{subsec:Ti_origin}

Thanks to the unprecedented energy resolution of Resolve and the rich photon statistics of Xtend data,
we make precise measurements of the abundances of IGEs in the SE and NE regions.
The Resolve spectrum revealed clear differences in spectral line features between the IMEs and IGEs:
IME lines are narrow, whereas IGE lines are broad.
In particular, the IGE lines show non-Gaussian profiles, suggesting either
bulk line-of-sight motion or a superposition of multiple plasma components along the line of sight.
Our spectral analysis indicates that the IGE lines are well reproduced by two NEI components
sharing common abundance ratios and velocity dispersions but having different electron temperatures
and ionization timescales.
The observed Cr/Fe and Ti/Fe ratios in the SE region are higher than those in the adjacent NE region,
indicating a localized enhancement of Cr and Ti.
Although the detection of Ti has been reported with the XMM-Newton observation \citep{Ohshiro21},
the statistical uncertainties of Ti/Fe and Cr/Fe are reduced from $\sim 30\%$ to $\sim 15\%$.
In this subsection, we constrain the progenitor central density using the same method as \citet{Ohshiro21}.

According to the nucleosynthesis models of SN Ia, IGEs are produced mainly in three different regimes
\citep{Bravo13, Leung18}: incomplete Si-burning, NSE, and n-rich NSE.
Matter in the n-rich NSE undergoes efficient electron capture and shifts nucleosynthesis yields towards
neutron-rich abundances, i.e., $Y_e < 0.5$ \citep{Lach20, Kanji20}, where $Y_e$
is the mean number of electrons per baryon.
Therefore, the abundance ratios of IGEs provide some of the strongest diagnostics of the central density \citep{Leung18}.
Especially, the density of the ejecta when the temperature of the WD becomes the maximum value
determines the degree of neutronization and subsequently determines mass ratios among Ti and IGEs.

Before we compare the observation with theoretical models,
we estimate the amount of iron contained in the SE region using the flux of K-shell emission lines.
The flux $F$ of the iron K-shell emission line can be expressed as
\begin{equation}
    \label{eq:iron_flux}
    F = \int \frac{\varepsilon_{\mathrm{FeK}\alpha} (kT_{\mathrm{e}})}{4\pi d^2} n_{\mathrm{e}} n_{\mathrm{Fe}}\, d\mathrm{V},
\end{equation}
\noindent
where $\varepsilon_{\mathrm{FeK}\alpha} (kT_{\mathrm{e}})$ is the emissivity of the K-shell lines of iron, $d$ is the distance to the source, $n_{\mathrm{e}}$ and $n_{\mathrm{Fe}}$ are the electron and iron number densities, respectively, and $d\mathrm{V}$ is the volume element.
We consider that the iron, with a mass $M_{\mathrm{Fe}}$, which emits the K-shell lines, is uniformly distributed within a volume $V$ and observed within a solid angle $\Omega$.
When we assumed that $V \propto d^3 \Omega^{3/2}$, $n_{\mathrm{Fe}} \propto M_{\mathrm{Fe}}/V$, and $n_{\mathrm{e}} \propto n_{\mathrm{Fe}}$ (i.e., free electrons are supplied by the ionization of iron), $M_{\mathrm{Fe}} \propto \Omega^{3/4} F^{1/2}$ can be derived.
Therefore, the iron mass ratio of the SE region $M_{\mathrm{Fe, SE}}$ to the entire region $M_{\mathrm{Fe, Total}}$ can be estimated as
\begin{equation}
    \label{eq:iron_mass_ratio}
    \frac{M_{\mathrm{Fe, SE}}}{M_{\mathrm{Fe, Total}}} = \left( \frac{F_{\mathrm{SE}}}{F_{\mathrm{Total}}} \right)^{1/2} \left( \frac{\Omega_{\mathrm{SE}}}{\Omega_{\mathrm{Total}}} \right)^{3/4}.
\end{equation}
We measured the flux of the iron K-shell emission lines from both the SE region and the entire SNR
using the Xtend data because the Resolve FoV does not cover the whole SNR.
In this observation, we obtained
$F_{\mathrm{SE}} = 1.5 \times 10^{-12}\ \mathrm{erg~s^{-1}~cm^{-2}}$, $F_{\mathrm{Total}} = 1.7 \times 10^{-12}\ \mathrm{erg~s^{-1}~cm^{-2}}$, $\Omega_{\mathrm{SE}} = 4.8 \times 10^{-7}\ \mathrm{sr}$,
and $\Omega_{\mathrm{Total}} = 1.8 \times 10^{-6}\ \mathrm{sr}$,
resulting in $M_{\mathrm{Fe, SE}}/M_{\mathrm{Fe, Total}} = 0.20$.
This result indicates that the SE region contains $\sim 0.16\ M_{\odot}$ of iron,
assuming a total Fe mass of $\sim 0.8\ M_{\odot}$ expected from typical near-$M_{\mathrm{Ch}}$ SNe~Ia events.

\begin{figure*}[tbp]
 \begin{center}
  \includegraphics[width=\textwidth]{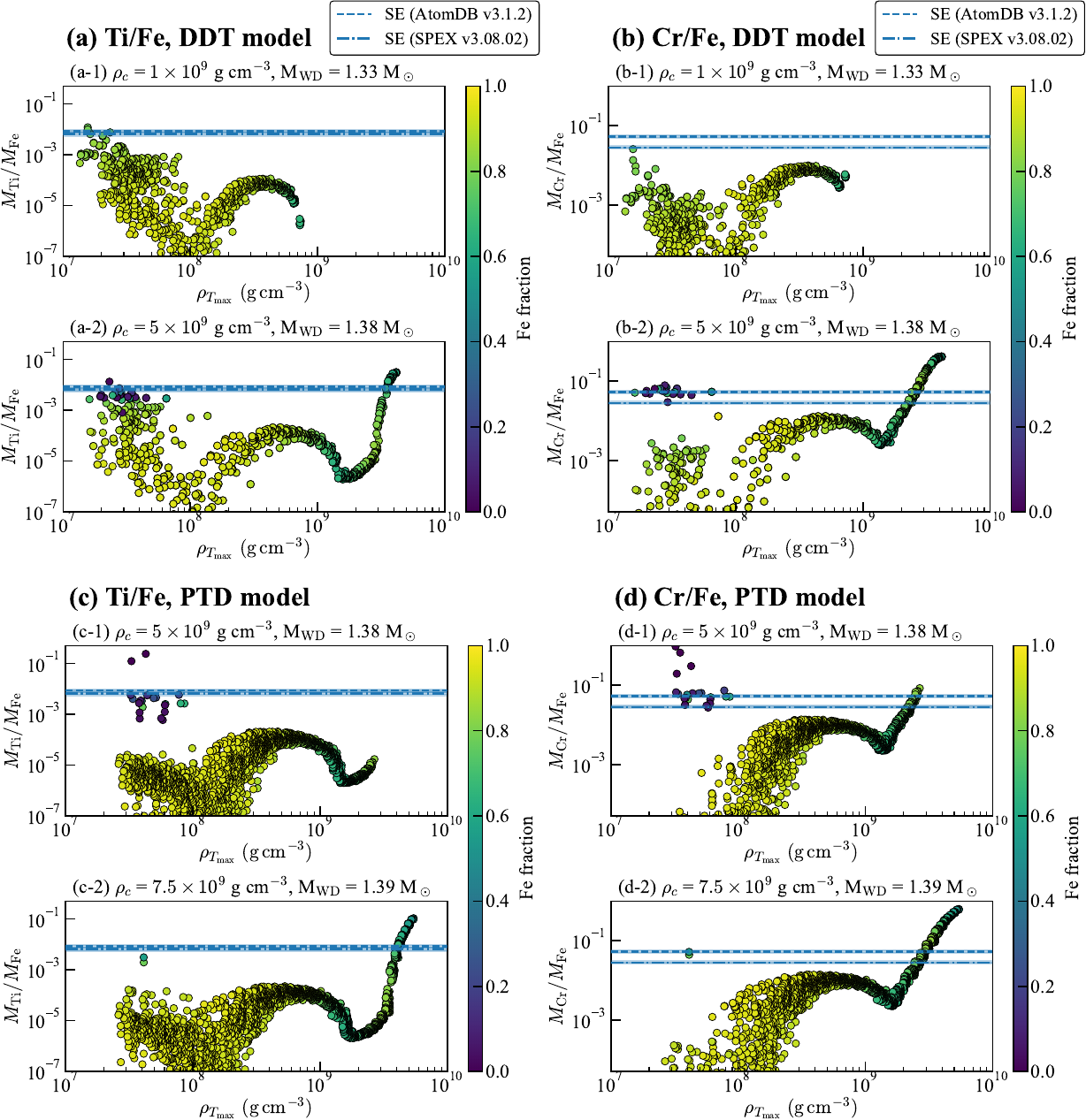} 
 \end{center}
\caption{
    Predicted distributions of the Ti/Fe and Cr/Fe mass ratios as a function of $\rho_{T_{\rm max}}$, the density at which each tracer particle reaches its maximum temperature during nuclear burning, for near-$M_{\rm Ch}$ SN~Ia models with solar metallicity in \citet{Leung18} and \citet{Leung20}. 
    (a) Ti/Fe distribution predicted by the deflagration-to-detonation transition (DDT) models
    for $\rho_c=1\times10^9$~g~cm$^{-3}$ (a-1) and $\rho_c=5\times10^9$~g~cm$^{-3}$ (a-2).
    (b) is the same as (a) but for Cr/Fe.
    (c) and (d) are the same as (a) and (b), respectively, but for the pure turbulent deflagration (PTD) models
    with $\rho_c=5\times10^9$~g~cm$^{-3}$ (\nobreakdash-$1$) and $\rho_c=7.5\times10^9$~g~cm$^{-3}$ (\nobreakdash-$2$).
    Only tracer particles with $T_{\rm max}\ge 5.5\times10^9$~K are shown, and the Fe mass fraction of each particle is color-coded.
    The observed mass ratios in the SE region, derived with AtomDB (dashed) and SPEX (dash-dot), are overlaid as horizontal bands indicating the $1\sigma$ statistical uncertainty.
{Alt text: Eight line graphs showing the comparison between the calculated mass ratios of Ti/Fe and Cr/Fe and measured values
as a function of $\rho_{T_{\rm max}}$.}    }\label{fig:model_mass_distribution}
\end{figure*}

Then we compare the observed mass ratios with theoretical nucleosynthesis models based on
two-dimensional hydrodynamic simulations of a near-$M_{\mathrm{Ch}}$ WD with solar metallicity,
calculated for different central densities ($\rho_{c}$) and explosion mechanisms.
Two explosion mechanisms are considered: a deflagration-to-detonation transition
(DDT; \cite{Khokhlov91, Poludnenko19}), with central ignition of the carbon
deflagration and a detonation-transition density of $\sim 2 \times 10^{7}\ \mathrm{g~cm^{-3}}$,
and a pure turbulent deflagration (PTD; \cite{Nomoto76,Gamezo2004,Fink14}),
in which the flame remains a deflagration and never transitions to a detonation. 
Figures~\ref{fig:model_mass_distribution}(a\nobreakdash-1) and \ref{fig:model_mass_distribution}(a\nobreakdash-2)
show Ti/Fe distributions predicted by the DDT models as a function of $\rho_{T_{\mathrm{max}}}$,
the density at which each tracer particle reaches its maximum temperature during nuclear burning,
for $\rho_{c}=1 \times 10^{9}\ \mathrm{g~cm^{-3}}$ (low density)
and $\rho_{c}=5 \times 10^{9}\ \mathrm{g~cm^{-3}}$ (high density), respectively;
Figures~\ref{fig:model_mass_distribution}(b\nobreakdash-1) and \ref{fig:model_mass_distribution}(b\nobreakdash-2)
present the corresponding Cr/Fe distributions for the same low- and high-density cases.
Figures~\ref{fig:model_mass_distribution}(c\nobreakdash-1)
and \ref{fig:model_mass_distribution}(c\nobreakdash-2) present the analogous
Ti/Fe distributions for the PTD models with
$\rho_{c}=5 \times 10^{9}\ \mathrm{g~cm^{-3}}$ and $\rho_{c}=7.5 \times 10^{9}\ \mathrm{g\,cm^{-3}}$,
respectively, and Figures~\ref{fig:model_mass_distribution}(d\nobreakdash-1)
and \ref{fig:model_mass_distribution}(d\nobreakdash-2) show the corresponding Cr/Fe distributions.
Regardless of the explosion mechanism, only the innermost ejecta in the high-density case simultaneously
reproduce the observed mass ratios of $M_{\mathrm{Ti}}/M_{\mathrm{Fe}} \sim 0.005$
and $M_{\mathrm{Cr}}/M_{\mathrm{Fe}} \sim 0.05$.

\begin{figure*}[tbp]
 \begin{center}
  \includegraphics[width=\textwidth]{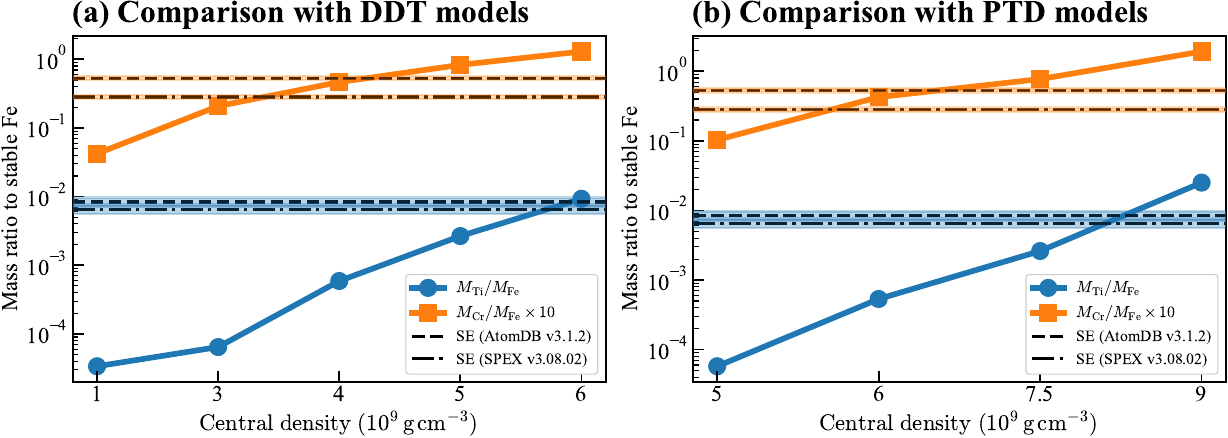} 
 \end{center}
\caption{
    Predicted mass ratios of Ti/Fe (blue circles) and Cr/Fe (orange squares) in the innermost ejecta (see text) as a function of the WD central density.
    Panel (a) shows DDT models, and panel (b) shows PTD models.
    The observed ratios in the SE region derived with AtomDB (dashed) and SPEX (dash-dot) are overlaid as horizontal bands indicating the $1\sigma$ statistical uncertainty.
    The Cr/Fe ratios are multiplied by 10 for clarity.
{Alt text: Two line graphs showing the comparison between the calculated mass ratios of Ti/Fe and Cr/Fe and measured values
as a function of the central density.}    }\label{fig:constrain_rhoc}
\end{figure*}

In Figures~\ref{fig:constrain_rhoc}(a) and \ref{fig:constrain_rhoc}(b), we show the Ti/Fe and Cr/Fe
mass ratios predicted for the innermost ejecta as a function of the central density, $\rho_{c}$,
for the DDT and PTD models, respectively.
Here we define the ``innermost ejecta'' as the highest-$\rho_{T_{\mathrm{max}}}$ region whose
cumulative Fe mass reaches 20\% of the total Fe yield (i.e., $M_{\mathrm{Fe}} \simeq 0.16\,M_{\odot}$
for the total Fe yield of $0.8 M_{\odot}$), which enables a direct comparison with the observed
mass ratios of the 3C~397 southeastern ejecta clump.
For the DDT case, the observed Ti/Fe and Cr/Fe ratios require
$\rho_{c} \gtrsim 5 \times 10^{9}\ \mathrm{g\,cm^{-3}}$
and $\rho_{c} \gtrsim 3 \times 10^{9}\ \mathrm{g\,cm^{-3}}$,
respectively, consistent with the previous XMM--Newton result \citep{Ohshiro21}.
For the PTD case, even higher central densities are required,
namely $\rho_{c} \gtrsim 7.5 \times 10^{9}\ \mathrm{g\,cm^{-3}}$
for Ti/Fe and $\rho_{c} \gtrsim 5 \times 10^{9}\ \mathrm{g\,cm^{-3}}$ for Cr/Fe.

\subsection{Localization of the ejecta}


The off-center localization of Ti and Cr is thought-provoking because
these elements are expected to trace neutronized ashes produced
by electron captures at high density in the innermost layers of the WD.
In near-$M_{\rm Ch}$ explosion scenarios, after a thermonuclear runaway
is triggered near the center, the burning is generally expected to start
as a subsonic deflagration flame.
The hot ashes produced near the center become buoyant relative to
the surrounding unburned material, rising outward as plumes driven
by the Rayleigh--Taylor instability.
Multidimensional simulations of SNe~Ia have demonstrated that such
ashes can be transported to the outer layers of the ejecta
(e.g., \cite{Maeda2010,Seitenzahl13}).
In particular, \citet{Mehta24} examined two-dimensional
hydrodynamical explosion models motivated by 3C~397 and showed that
PTD models naturally place neutronized species in clumpy structures
in the outer ejecta, whereas the DDT model retains a more stratified
ejecta structure with neutronized material concentrated near the center.
Thus, the observed off-center distributions of Ti and Cr may indicate
an explosion in which the deflagration phase persisted unusually
long before detonation, or one in which no detonation occurred.
Such a scenario is not ruled out by current theory,
because the conditions leading to DDT
remain uncertain and have long been modeled using
prescribed criteria in SNe~Ia simulations
(e.g., \cite{Iwamoto99,ciaraldi-schoolmann2013}).
Although a physically motivated criterion for triggering the transition
has been proposed \citep{poludnenko2019} and recently
implemented in SNe~Ia simulations \citep{patel2026},
the parameter space in which the deflagration phase either persists
until late times or never triggers detonation has not yet been
fully explored within this framework.
The trigger of detonation primarily depends on the diffusion strength
of the turbulent deflagration front to reach the distributed burning
\citep{Brooker2021}, but it is unclear if such criteria can be
unconditionally satisfied in an arbitrary flame kernel.

From an observational perspective, distinguishing between DDT and PTD
explosions is crucial, as the two scenarios imply substantially
different SN properties.
The former are widely invoked for normal SNe~Ia
(e.g., \cite{kasen2009,blondin2011,sim2013}),
whereas the latter have been discussed in relation to
SN2002cx-like or Type~Iax SNe (e.g., \cite{Li03,Foley13,Jha17}),
which show lower ejecta velocities and lower luminosities than
normal SNe~Ia.
A key observational discriminant between these scenarios is
the amount of unburned material: a PTD explosion
(i.e., no detonation) can leave a substantial amount of
unburned C and O (e.g., \cite{Fink14,Kromer15,Leung20}),
whereas a DDT explosion more efficiently burns the C and O
left after the initial deflagration (e.g., \cite{Gamezo2004,Leung18}).
Detecting shocked C and O ejecta could therefore provide
a useful diagnostic of whether detonation occurred.
Such material may appear not only through C and O emission lines
but also through its contribution to the broadband X-ray continuum,
as illustrated by the Tycho SNR modeling by \citet{badenes2006}.
For 3C~397, however, this diagnostic is not straightforward
to apply because the large interstellar absorption suppresses
the soft X-ray emission from C and O.
A definitive test of the PTD scenario would require full SNR modeling
that self-consistently treats the explosion ejecta structure,
shock evolution, ionization state, and broadband X-ray spectrum,
which is beyond the scope of this work.

\subsection{Ni/Fe mass ratio}

\begin{figure}[tbp]
 \begin{center}
  \includegraphics[width=\columnwidth]{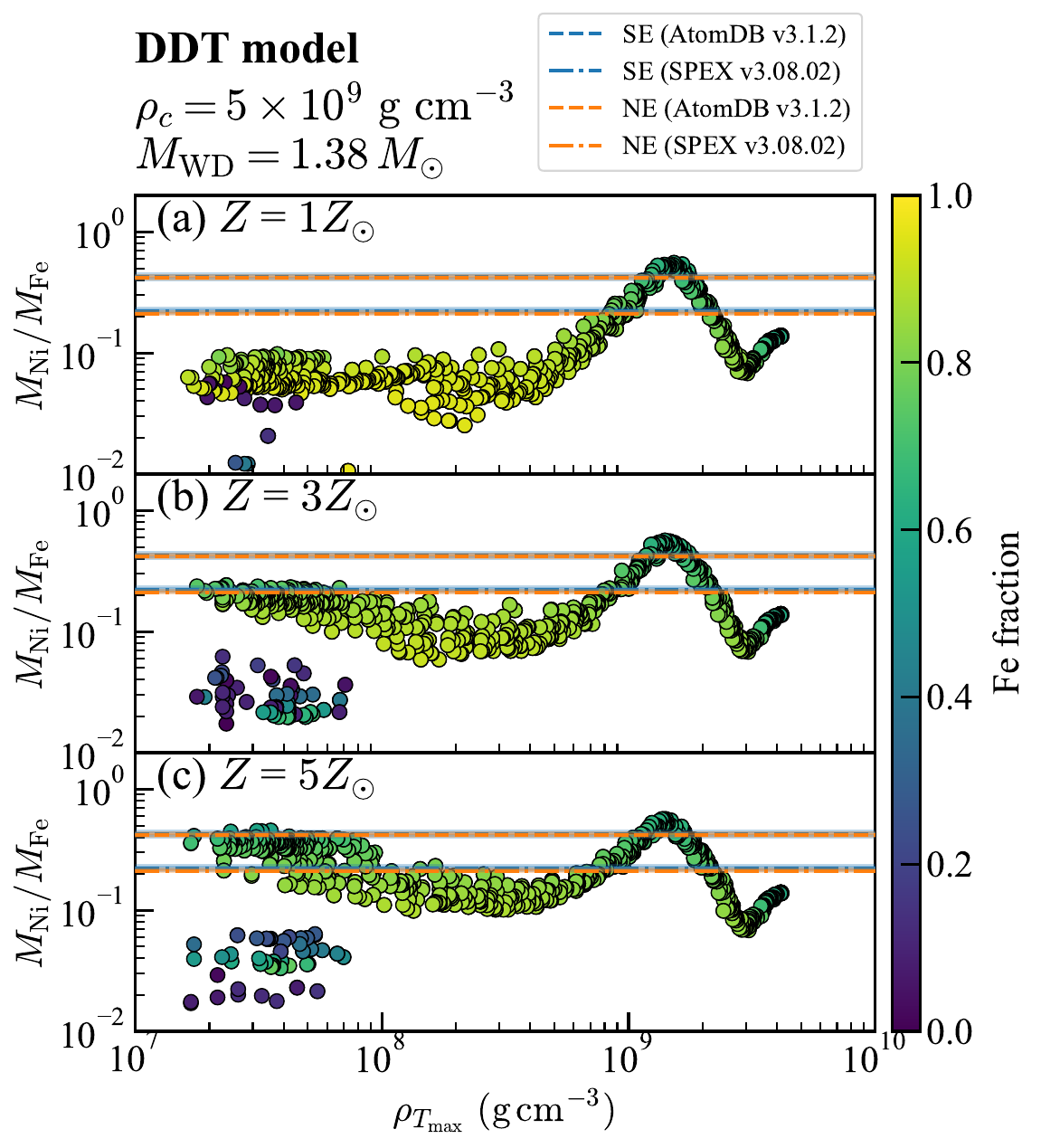} 
 \end{center}
\caption{
    Same as Fig.~\ref{fig:model_mass_distribution}, but for the Ni/Fe mass ratio predicted by a delayed-detonation (DDT) model with $\rho_c=5\times10^9$~g~cm$^{-3}$ at three progenitor metallicities, $Z=1,\,3,$ and $5\,Z_{\odot}$.
    The observed ratios in the SE (blue) and NE (orange) regions, derived using AtomDB (dashed) and SPEX (dash--dot), are overlaid as horizontal bands indicating the $1\sigma$ statistical uncertainty.
{Alt text: Three line graphs showing the comparison between the calculated mass ratio of Ni/Fe and measured values at three
different progenitor metallicities
as a function of $\rho_{T_{\rm max}}$.} }\label{fig:Ni_Fe_ratio}
\end{figure}

The mass ratio of Ni/Fe, measured with the highest precision ever,
is another strong probe of the progenitor properties.
Figure~\ref{fig:Ni_Fe_ratio} shows the mass ratio of Ni to Fe as a function
of $\rho_{T_{\rm max}}$, changing the metallicity of the progenitor for the DDT model.
Comparing the panels makes it clear that the difference of the progenitor metallicity
appears as the difference of the mass ratios at the lower density range of
$\rho_{T_{\rm max}}< 10^9~{\rm g~cm^{-3}}$.
This is because the origin of the neutronized species is different between the innermost
n-rich NSE and the outer layers; neutron excess in the outer layer depends
on the abundance of $^{22}$Ne produced via CNO cycle \citep{Timmes03},
while the neutron excess in the deepest core is dominated by electron capture
\citep{Brachwitz00}.
As shown in figure~\ref{fig:Ni_Fe_ratio}(a), observed values match the calculations only at the high density
range ($\rho_{T_{\rm max}} > 10^9~{\rm g~cm^{-3}}$) for both regions in the case of the solar metallicity.
However, assuming the similar density of the tracer particles between the two regions
is in tension with the implication from the observed spectra; the ejecta clump
in the SE region originates from the inner dense layer than that in the NE region.
On the other hand, higher-metallicity cases indicate that
a density range lower than $10^8~{\rm g~cm^{-3}}$ is also preferable, as shown
in figure~\ref{fig:Ni_Fe_ratio}(b) and \ref{fig:Ni_Fe_ratio}(c).
These cases are compatible with the view that only the ejecta in the SE region comes
from the innermost layer.
In particular, if the five times of the solar value is assumed, the model explains
the observational results regardless of the atomic databases in the low density range.
To discriminate the progenitor metallicity between the latter two cases, the difference
due to the atomic database is crucial. As found by \citet{Plucinsky25} in the spectral
fitting of Cassiopeia A that primarily composed of thermal plasma in NEI state,
Ni abundance differs by a factor of two between \texttt{SPEX} and \texttt{AtomDB}.
We observe roughly the same degree of difference in Ni abundances between the databases.
This might be due to uncertainties in the strengths of the satellite lines
of Li, Be, and B-like Fe ions slightly above 8~keV. To investigate the progenitor metallicity
with more precision, resolving this discrepancy is awaited.



\subsection{Ionization degrees of IMEs and IGEs}

Another noteworthy feature in the XRISM spectral analysis is the unusually high ionization state of the ejecta.
The IME components have ionization timescales of $\tau \gtrsim 10^{12}{\rm cm^{-3}s}$, indicating that they are close to collisional ionization equilibrium \citep{Smith10}, whereas Ti and the IGEs remain in an ionizing state with $\tau \sim 10^{10}$–$10^{11}{\rm cm^{-3}s}$.
This difference is qualitatively expected for stratified ejecta: IMEs synthesized in the outer layers are encountered by the reverse shock earlier than IGEs synthesized deeper inside, giving them longer post-shock timescales and hence larger values of $n_{\rm e}t$ \citep{Badenes07,Kosenko10}.

The key point for 3C~397 is that its inferred ionization state is higher than those measured in many other Type~Ia SNRs \citep{yamaguchi2014}, suggesting that stratification alone is insufficient to explain the observed ionization structure.
A natural way to increase the ionization timescale is to shock the ejecta early in a dense environment, so that the shocked plasma can subsequently undergo substantial ionization evolution.
This interpretation is supported by \citet{Court26}, who showed, through hydrodynamical simulations coupled with NEI spectral calculations, that the radius, age, and Fe-K$\alpha$ properties of 3C~397 favor interaction with material denser than typical interstellar medium densities, pointing to strong interaction with circumstellar material (CSM).
Such dense or structured material may have an observational counterpart in the CO data, which indicates that 3C~397 is associated with a cavity-like molecular shell possibly formed before the supernova explosion \citep{Ito25}.

A similar situation is present in N103B, which is also considered
a candidate for a single-degenerate, near-$M_{\rm Ch}$ explosion
(e.g., \cite{Yamaguchi21}).
Indeed, within the framework proposed by \citet{Court26},
these properties can be interpreted as a natural consequence
of strong interaction with the CSM, placing N103B in the same class
as 3C~397.
This interpretation is also supported by the cavity-like molecular
cloud reported around N103B \citep{Sano18}.
These similarities suggest that dense or structured surrounding
material may be relevant to the ionization evolution of
candidate single-degenerate, near-$M_{\rm Ch}$ Type~Ia SNRs.

\section{Summary}

We have reported the first result from the 3C~397 observation performed during
the PV phase of the XRISM. The FoV of Resolve is set to the eastern lobe of
the remnant, which contains the Cr-rich region identified by the past
observation.
High-resolution spectra by Resolve clearly show the narrow emission lines
from IMEs, enabling us to measure the plasma temperature more precisely
than in past observations.
Extraordinarily strong emission lines from IGEs are
detected throughout the observed region. The line intensities and profiles
from Ti and Cr exhibit a clear difference between northern and southern
parts of the data, which confirms the localization of the neutronized ejecta
into the off-center position. Ni abundance is precisely measured using
statistically rich spectra obtained by Resolve and Xtend.
We fit the spectra using the NEI plasma model with the line broadening.
Assessing the contamination from the nearby point source and NXB, we
evaluate the mass ratios among Ti and IGEs. The latest versions of 
\texttt{SPEX} and \texttt{AtomDB} are adopted to estimate the systematic
uncertainties of the fitted parameters.
We have successfully constrained the central density of the progenitor
by comparing the observational data to the multidimensional calculation
for both of the DDT and PTD models. The derived range of the density
is $\gtrsim (3-5) \times 10^{9}\ \mathrm{g\,cm^{-3}}$ for the DDT
and $\gtrsim 5-7.5 \times 10^{9}\ \mathrm{g\,cm^{-3}}$ for the PTD case.
Both cases support the quite high densities that require
$M_{\rm Ch}$ explosion.

The off-center localization of Ti and Cr can be interpreted as the result of buoyancy-driven outward transport of neutronized ashes synthesized near the WD center during the deflagration phase.
Such a configuration may arise if the deflagration phase persisted unusually long before detonation, or if no detonation occurred.
Mass ratio of Ni and Fe would put a constraint on the progenitor
metallicities with 3-5 solar values.

\begin{ack}
We would like to express our sincere gratitude to
all members of the scientists, engineers, and technicians
who realized the XRISM satellite. We also thank those who
operate the mission, and have developed the XRISM software.
This work was partly supported by Japan Society for the Promotion
of Science (JSPS) Grants-in-Aid for Scientific Research (KAKENHI)
Grant Numbers, JP24H00253 (H.N.), JP22KJ1047 (Y.O.), JP23K20850 (K.M), JP24K17105 (Y.K.),
22H00158, 23H01211 (H.Y.), 24K00677 (M.N.).
SCL acknowledges support from the National Science Foundation under Grant AST-2316807.
Y.O. acknowledges support from the Special Postdoctoral Researchers Program in RIKEN.
This work was supported by the JSPS Core-to-Core Program under Grant Number JPJSCCA20220002.
\end{ack}


\bibliographystyle{aasjournal}   
\bibliography{mybibfile}

\end{document}